# Curl-free positive definite form of time-harmonic Maxwell's equations well-suitable for iterative numerical solving

V.E. Moiseenko[1] and O. Ågren[2]

[1] *Institute of Plasma Physics, National Science Center "Kharkiv Institute of Physics and Technology", 61108 Kharkiv, Ukraine*

[2] *Uppsala University, S-75108 Uppsala, Sweden*

**Abstract**

A new form of time-harmonic Maxwell's equations is developed and proposed for numerical modeling. It is written for the magnetic field strength **H**, electric displacement **D**, vector potential **A** and the scalar potential $\Phi$. There are several attractive features of this form. The first one is that the differential operator acting on these quantities is positive. The second is absence of *curl* operators among the leading order differential operators. The Laplacian stands for leading order operator in the equations for **H**, **A** and $\Phi$, while the gradient of divergence stands for **D**. The third feature is absence of space varied coefficients in the leading order differential operators that provides diagonal domination of the resulting matrix of the discretized equations. A simple example is given to demonstrate the applicability of this new form of time-harmonic Maxwell's equations.

## 1. Introduction

The conventional form of time-harmonic Maxwell's equations in non-magnetic media is

$$\nabla \times \mathbf{E} = ik_0 \mathbf{H} \; ; \; \nabla \times \mathbf{H} = -ik_0 \hat{\boldsymbol{\varepsilon}} \cdot \mathbf{E} + \frac{4\pi}{c} \mathbf{j}_{ext}. \qquad (1\,a,b)$$

Here **E** and **H** are electric and magnetic field strengths, $\hat{\boldsymbol{\varepsilon}}$ is the dielectric tensor, $\mathbf{j}_{ext}$ is the external (antenna) current density, $k_0 = \omega/c$, $c$ is the speed of light and the time dependence of all quantities is realized through the multiplier $\exp(-i\omega t)$. With proper boundary conditions, such equations are relevant for problems of radio-frequency wave excitation and propagation in plasmas and other media. For finite element analysis, the first order equations (1 a,b) are often combined into the form

$$\nabla \times \nabla \times \mathbf{E} = k_0^2 \hat{\boldsymbol{\varepsilon}} \cdot \mathbf{E} + \frac{4\pi i \omega}{c^2} \mathbf{j}_{ext}, \qquad (2)$$

in which the leading derivatives are of second order.

Numerical solution of the boundary problem normally implies the discretization in space. The applied standard discretization techniques born spurious modes, both for Eqs. (1a,b) as well as

Eq. (2). Since $\nabla \cdot \nabla \times \mathbf{F} = 0$ (for any arbitrary vector $\mathbf{F}$), a set of the components of the *curl* operator is degenerate. The degeneration is cancelled by the dicretization procedure, and a new wave-like mode appears which has no physical nature. To avoid this, a number of methods have been proposed. The first one is the Yee template for the finite differences [1]. In this template the mesh is staggered, and every $\mathbf{E}$- and $\mathbf{H}$- field component occupies different mesh nodes. The different mesh modes introduce difficulties for imposing the boundary conditions, calculation of dissipated power, and reproducing correctly rotating electric field components, $E_\pm = E_x \pm E_y$, which are responsible for the cyclotron wave dumping in magnetized plasma (the steady magnetic field is in *z* direction). For Eq. (2) a special template has been proposed [2] and used in [3,4] with the same mesh for all components of the electric field. These two templates produce degenerate systems, and the spurious modes do not appear.

The Yee approach has a finite element analog, a Gruber-Rappaz method [5]. This method uses different order Lagrange and Hermit finite elements for different components of the electric field. Later on, special *curl*-conforming finite elements were invented like Nedelec finite elements [6]. Maxwell's equations in an alternative form can be obtained by introducing the vector and scalar potentials substituting $\mathbf{E} = ik_0\mathbf{a} - \nabla\varphi$ and $\mathbf{H} = \nabla \times \mathbf{a}$ yield with the Coulomb gauge $\nabla \cdot \mathbf{a} = 0$

$$\Delta\mathbf{a} + k_0^2 \hat{\boldsymbol{\varepsilon}} \cdot \mathbf{a} + ik_0 \hat{\boldsymbol{\varepsilon}} \cdot \nabla\varphi = -\frac{4\pi}{c}\mathbf{j}_{ext} \tag{3}$$

$$\nabla \cdot [\hat{\boldsymbol{\varepsilon}} \cdot (ik_0\mathbf{a} - \nabla\varphi)] = 4\pi\rho_{ext} \tag{4}$$

Here $\rho_{ext}$ is the electric charge. Here and further on, the Laplacian of a vector is understood as $\Delta\mathbf{F} = \nabla\nabla \cdot \mathbf{F} - \nabla \times \nabla \times \mathbf{F}$. The form (3-4) has no degenerate operators and can be dicretized in a standard way which is a serious benefit.

After discretization a system of linear algebraic equations $\mathbf{Mx}=\mathbf{b}$ appears that could be solved by direct or iterative methods. 1D problem produces a dense band matrix and is best for direct methods, while direct and iterative methods compete in 2D problems, and 3D is for iterative methods.

All the above forms of the Maxwell's equations are sign indefinite and produce sign indefinite matrices, i.e. the scalar product $(\mathbf{f}, \mathbf{Mf})$ could have any value ( here $\mathbf{f}$ is an arbitrary vector of the domain).

For sign indefinite linear problems we cannot apply iterative methods of relaxation family and the conjugate gradients methods which otherwise offer good convergence rates. The GMRes, BiCGStab [7] and other methods are developed for this. The iterations for Maxwell's equations do not show good convergence and often stagnate [8]. The inner and outer iterations are used to avoid this [9], but performance of the iterations is still low. Expensive, but the robust approach is to retain time dependence in Maxwell's equations while the problem is essentially time-harmonic [10].

The aim of the paper is to develop sign definite form of Maxwell's equations, discretization of which likely results in sign definite matrix for linear equations and allow one to use efficient iterations to obtain the numerical solutions. Another aim is to eliminate the *curl* operator at leading derivatives. This facilitates and simplifies the discretization and likely boosts the iterations.

## 2. New alternative form of Maxwell's equations

The source equations for this study are basically equations (1) written in the following form

$$\nabla \times \mathbf{A} - ik_0 \mathbf{H} = 0, \tag{5}$$

$$\nabla \times \mathbf{H} + ik_0 \mathbf{D} = \frac{4\pi}{c} \mathbf{j}_{ext}, \tag{6}$$

$$\nabla \cdot \mathbf{D} = 4\pi \rho_{ext}, \tag{7}$$

$$\nabla \Phi + ik_0 \hat{\boldsymbol{\zeta}} \cdot \mathbf{D} - ik_0 \mathbf{A} = 0, \tag{8}$$

Here $\mathbf{D}$ is the electric displacement, and $\mathbf{A} = ik_0 \mathbf{a}$, $\Phi = -i\varphi/k_0$ are normalized potentials. The introduction of these normalized potentials is made to have all quantities in the equations (5-7) of the same order of magnitude. This is done having in mind further discretization and combination of all discretized values to the single vector $\mathbf{x}$ of the linear system. The material equation is $\hat{\boldsymbol{\zeta}} \cdot \mathbf{D} = \mathbf{E}$ instead of the commonly used $\mathbf{D} = \hat{\boldsymbol{\varepsilon}} \cdot \mathbf{E}$. This change is motivated by the want to have no coefficients before the differential operators of a leading order in (5-8).

With charge continuity assumed to be met, equations (5-7) are dependent since Eq. (7) can be obtained from Eq. (5). This should be kept in mind during the further transforms.

Let us denote the left hand sides of the Eqs. (5-8) as $\mathbf{L}_A$, $\mathbf{L}_H$, $L_D$ and $\mathbf{L}_\Phi$, and right hand sides as $\mathbf{R}_H$ and $R_D$. Then we construct the following quadratic form

$$Q = \int dV [\mathbf{L}_A^* \cdot \mathbf{L}_A + \mathbf{L}_H^* \cdot (\mathbf{L}_H - \mathbf{R}_H) + L_D^* (L_D - R_D) + \mathbf{L}_\Phi^* \cdot \mathbf{L}_\Phi]. \tag{9}$$

This form is zero valued if the operators $\mathbf{L}$ act on the solution of the system (5-7). $Q$ is real and positive if $\mathbf{R}_H = 0$ and $R_D = 0$, and $\mathbf{A}$, $\mathbf{H}$, $\mathbf{D}$ and $\Phi$ are arbitrary. Using Gauss's theorem, $Q$ is transformed into

$$Q = Q_S + Q_V. \tag{10}$$

The surface term reads

$$Q_S = \int d\mathbf{S} \cdot (\mathbf{A}^* \times \mathbf{L}_A) + \int d\mathbf{S} \cdot [\mathbf{H}^* \times (\mathbf{L}_H - \mathbf{R}_H)] + \int dS D^* (L_D - R_D) + \int \Phi^* d\mathbf{S} \cdot \mathbf{L}_\Phi, \tag{11}$$

where $d\mathbf{S}$ is the surface element of the surface surrounding the domain and directed outward. The volume integral part is

$$Q_V = \int dV[\mathbf{A}^* \cdot \mathbf{L}'_A + \mathbf{H}^* \cdot (\mathbf{L}'_H - \mathbf{R}_H) + \mathbf{D}^* \cdot (\mathbf{L}'_D - \mathbf{R}'_D) + \Phi^* L'_\Phi] \qquad (12)$$

with $\mathbf{L}'_A = \nabla \times \mathbf{L}_A + ik_0 \mathbf{L}_\Phi$, $\mathbf{L}'_H = ik_0 \mathbf{L}_A + \nabla \times \mathbf{L}_H$, $\mathbf{R}'_H = \nabla \times \mathbf{R}_H$,

$\mathbf{L}'_D = -ik_0 \mathbf{L}_H - \nabla L_D - ik_0 \hat{\zeta}^{T*} \cdot \mathbf{L}_\Phi$, $\mathbf{R}'_D = -ik_0 \mathbf{R}_H - \nabla R_D$, $L'_\Phi = -\nabla \cdot \mathbf{L}_\Phi$

If $\mathbf{A}$, $\mathbf{H}$, $\mathbf{D}$ and $\Phi$ are the solutions of the system (5-8) $Q_S$ is zero, while for arbitrary vectors $\mathbf{A}$, $\mathbf{H}$, $\mathbf{D}$ and $\Phi$, $Q_S$ could have any value. For further consideration we narrow the space of the arbitrary vectors which are considered in the problem of sign definition to those one for which each of four terms in (11) nullify (with $\mathbf{R}_H=0$ and $R_D=0$), i.e.

$$\int d\mathbf{S} \cdot (\mathbf{A}^* \times \mathbf{L}_A) = 0, \ \int d\mathbf{S} \cdot (\mathbf{H}^* \times \mathbf{L}_H) = 0, \ \int d\mathbf{S} D^* L_D = 0, \ \int \Phi^* d\mathbf{S} \cdot \mathbf{L}_\Phi = 0. \qquad (13)$$

However, it is assumed that the space remains rather wide to contain the vectors which can reproduce the solution with any prescribed accuracy.

We combine the quantities vectors $\mathbf{A}$, $\mathbf{H}$, $\mathbf{D}$ and $\Phi$ into one 10-component vector $\mathbf{X}$ and denote the corresponding 10-component operator by $\mathbf{L}$. With such a notation the volume integral is

$$Q_V = \int dV \mathbf{X}^* \cdot \mathbf{L}(\mathbf{X}) \qquad (14)$$

If the conditions (13) are met for vectors $\mathbf{X}$, $Q_V>0$. This means, that operator $\mathbf{L}$ is positive. The explicit form of the equations $\mathbf{L}(\mathbf{X})=\mathbf{R}$ (where $\mathbf{R}$ is the combined right-hand side) is

$$-\Delta \mathbf{A} + k_0^2 \mathbf{A} - ik_0 \nabla \times \mathbf{H} - k_0^2 \hat{\zeta} \cdot \mathbf{D} + ik_0 \nabla \Phi = 0, \qquad (15)$$

$$-\Delta \mathbf{H} + k_0^2 \mathbf{H} + ik_0 \nabla \times \mathbf{A} + ik_0 \nabla \times \mathbf{D} = \frac{4\pi}{c} \nabla \times \mathbf{j}_{ext}, \qquad (16)$$

$$-\nabla \nabla \cdot \mathbf{D} + k_0^2 (\mathbf{D} + \hat{\zeta}^{*T} \cdot \hat{\zeta} \cdot \mathbf{D}) - k_0^2 \hat{\zeta}^{*T} \cdot \mathbf{A} - ik_0 \nabla \times \mathbf{H} - ik_0 \hat{\zeta}^{*T} \cdot \nabla \Phi = -ik_0 \frac{4\pi}{c} \mathbf{j}_{ext} - 4\pi \nabla \rho_{ext}, \qquad (17)$$

$$-\Delta \Phi + ik_0 \nabla \cdot \mathbf{A} - ik_0 \nabla \cdot \hat{\zeta} \cdot \mathbf{D} = 0. \qquad (18)$$

Here $(\hat{\zeta}^{*T})_{ik} = (\hat{\zeta})^*_{ki}$. In obtaining the above system, the formulas $\nabla \times \nabla \times \mathbf{A} = -\Delta \mathbf{A}$ and $\nabla \times \nabla \times \mathbf{H} = -\Delta \mathbf{H}$ have been used which follow from the Coulomb gauge applied on $\mathbf{A}$

$$\nabla \cdot \mathbf{A} = 0 \qquad (19)$$

and from the divergence-free property of $\mathbf{H}$

$$\nabla \cdot \mathbf{H} = 0 \qquad (20)$$

which is a consequence of (5). In this way the *curl* operator is deleted from the leading derivatives in the system (15-18). In addition, after such a transformation the dependency of equations which is inherited from the system (5-8) is now cancelled.

The system (15-18) supports the divergence-free property of $\mathbf{A}$ and $\mathbf{H}$ in the following way

$$\Delta(\nabla \cdot \mathbf{A}) = 0, \tag{20}$$

$$\Delta(\nabla \cdot \mathbf{H}) - k_0^2 (\nabla \cdot \mathbf{H}) = 0. \tag{21}$$

Each of these equations obtained from (15-18) is isolated and have no right-hand side. They have zero valued solutions when the compatible boundary conditions are imposed to the system (15-18).

The requirement to the discretization applied to the system (15-18) is not only to reproduce the vector $\mathbf{X}$ by the vector of discrete values $\mathbf{x}$, the system of differential equations $\mathbf{L}(\mathbf{X})=\mathbf{R}$ by the linear system $\mathbf{Mx}=\mathbf{b}$, but also to reproduce the integral $I = \int dV \, \mathbf{X}^* \cdot \mathbf{Y}$ by the dot product

$$\tilde{I} = (\mathbf{x}, \mathbf{y}) = \sum_i x_i^* y_i.$$

## 3. The numerical exercise

### 3.1. Discretizetion procedure

As a simple example, the problem of electromagnetic field distribution in the metallic rectangular cavity (parallelepiped) is analyzed. The boundaries of the cavity are: $x=0$, $x=L_x$; $y=0$, $y=L_y$; $z=0$, $z=L_z$. The cavity is assumed to be empty with $\hat{\zeta}=1$. For the numerical solution we use the finite difference method. The coordinates are Cartesian, and the system (15-18) is projected to the coordinates. The mesh chosen is rectangular and equidistant. The differential operators are reproduced by the central differences.

In our numerical exercise the tangent current densities and the charge density nullifies at the boundary surface: $j_{ext,y}\big|_{x=0} = 0$, $j_{ext,z}\big|_{x=0} = 0$ and $\rho_{ext}\big|_{x=0} = 0$. This simplifies further calculations. As for the boundary conditions, we here consider application them at the surface $x=0$. On other surfaces, the boundary conditions are applied in an analogous way. On the surface $x=0$ the tangent components of $\mathbf{A}$ nullify $A_y\big|_{x=0} = 0$ and $A_z\big|_{x=0} = 0$. This turns to zero the first equation of (13) since $d\mathbf{S} \times \mathbf{A}^* = 0$. For calculation of $A_x$ the discretized $x$-component of equation (15) is used. The problem appears in discretization of $\dfrac{\partial^2}{\partial x^2} A_x$ for which only two points of mesh along the $x$-direction are available, $x_1=0$ and $x_2=h_x$, where $h_x$ is the mesh step in $x$ direction. Then we use equation (19) which gives $\dfrac{\partial}{\partial x} A_x \bigg|_{x=0} = 0$. With account for the latter,

$$\left.\frac{\partial^2}{\partial x^2} A_x\right|_{x=x_1, y=y_j, z=z_k} \to 2 \frac{-A_x^{(1,j,k)} + A_x^{(2,j,k)}}{h_x^2} \tag{22}$$

In this indirect way the divergence of **A** is set to zero at the boundary, and this provides $\nabla \cdot \mathbf{A} \approx 0$ in the domain as a solution of (20) (error due to the discretization may take place. but the boundary condition is handled).

As for the boundary conditions for the magnetic field, the condition $H_x|_{x=0} = 0$ can be found from the x-component of (5). Analyzing the tangent components of equation (6) one can find that $\frac{\partial}{\partial x} H_y\Big|_{x=0} = 0$ and $\frac{\partial}{\partial x} H_z\Big|_{x=0} = 0$. This allows one to use similar template for $\frac{\partial^2}{\partial x^2} H_{\{y \atop z\}}$ as in (22).

The boundary condition $H_x|_{x=0} = 0$ can be obtained directly from x-component of (16). But for that, one should assume that

$$\frac{\partial^2}{\partial x^2} H_x\Big|_{x=x_1} = 0. \tag{23}$$

Note here that $\frac{\partial^2}{\partial x^2} H_x\Big|_{x=x_1} = \frac{\partial}{\partial x} \nabla \cdot \mathbf{H}\Big|_{x=x_1}$. Formula (23) can be treated as the zero-valued Neumann boundary condition for equation (21). With such boundary conditions the equation has the solution $\nabla \cdot \mathbf{H} = 0$.

The second equation of (13) is nullified at the boundary since $d\mathbf{S} \times \mathbf{L}_H = 0$.

As for the electric displacement vector **D**, the boundary conditions are similar to those ones for **A**: $D_y|_{x=0} = 0$, $D_z|_{x=0} = 0$, and $\frac{\partial}{\partial x} D_x\Big|_{x=0} = 0$ is used to obtain the similar template for

$\frac{\partial^2}{\partial x^2} D_x\Big|_{x=x_1, y=y_j, z=z_k}$ as (22). The third equation of (13) nullify since $d\mathbf{S} \times \mathbf{D}^* = 0$.

And finally the electric potential is set to zero at the boundary nullifying the fourth equation of (13).

All the discretized equations are multiplied by the volume element. It is $V = h_x h_y h_z$ for the inner mesh nodes, twice smaller at the faces of the parallelepiped and four times smaller at the edges.

*3.2. Analytical solution*

To analyze the numerical results it is good to have an analytical solution for further comparison. The constructed analytical solution is the following.

The external current density is prescribed as

$$\begin{aligned}\mathbf{j}_{ext} &= \mathbf{e}_x j_{0x} \cos(k_x x)\sin(k_y y)\sin(k_z z) + \mathbf{e}_y j_{0y} \sin(k_x x)\cos(k_y y)\sin(k_z z) \\ &+ \mathbf{e}_z j_{0z} \sin(k_x x)\sin(k_y y)\cos(k_z z)\end{aligned} \tag{24}$$

where $k_x = \pi/L_x$, $k_y = \pi/L_y$, $k_z = \pi/L_z$. $j_{0x}$, $j_{0y}$ and $j_{0z}$ are constants. For such currents the analytical solutions are

$$\begin{aligned}\mathbf{A} = &\mathbf{e}_x A_{0x} \cos(k_x x)\sin(k_y y)\sin(k_z z) + \mathbf{e}_y A_{0y} \sin(k_x x)\cos(k_y y)\sin(k_z z) \\ &+ \mathbf{e}_z A_{0z} \sin(k_x x)\sin(k_y y)\cos(k_z z)\end{aligned} \quad (25)$$

with

$$\mathbf{A}_0 = \frac{4\pi i k_0}{c(k^2 - k_0^2)}\left[\mathbf{j}_0 - \frac{\mathbf{k}(\mathbf{k}\cdot\mathbf{j}_0)}{k^2}\right]. \quad (26)$$

Magnetic field expression is

$$\begin{aligned}\mathbf{H} = &\mathbf{e}_x H_{0x} \sin(k_x x)\cos(k_y y)\cos(k_z z) + \mathbf{e}_y H_{0y} \cos(k_x x)\sin(k_y y)\cos(k_z z) \\ &+ \mathbf{e}_z H_{0z} \cos(k_x x)\cos(k_y y)\sin(k_z z)\end{aligned} \quad (27)$$

and

$$\mathbf{H}_0 = (\nabla \times \mathbf{A}_0)/(ik_0). \quad (28)$$

The potential is

$$\Phi = \Phi_0 \sin(k_x x)\sin(k_y y)\sin(k_z z), \quad (29)$$

$$\Phi_0 = -\frac{4\pi}{ck_0^2}\mathbf{k}\cdot\mathbf{j}_0. \quad (30)$$

The electric displacement has a similar expression as (25) coefficients of which are calculated using the formula

$$\mathbf{D}_0 = \mathbf{A}_0 + i\mathbf{k}\Phi_0/k_0. \quad (31)$$

### 3.3. Numerical results

After discretization, the resulting linear system $\mathbf{Mx}=\mathbf{b}$ is solved using the Conjugate Gradients (CG) method. No matrix preconditioning is employed. In the calculations, the normalized residual

$$\eta_r^{(m)} = \sqrt{\frac{(\mathbf{Mx}^{(m)} - \mathbf{b}, \mathbf{Mx}^{(m)} - \mathbf{b})}{(\mathbf{b},\mathbf{b})}}. \quad (32)$$

Here $m$ enumerates iterations. The normalized error

$$\eta_e^{(m)} = \sqrt{\frac{(\mathbf{x}^{(m)} - \mathbf{x}_a, \mathbf{x}^{(m)} - \mathbf{x}_a)}{(\mathbf{x}_a,\mathbf{x}_a)}} \quad (33)$$

is also calculated ($\mathbf{x}_a$ contains the analytical solution values in the mesh nodes). The test show rather quick and smooth convergence (see Fig.1).

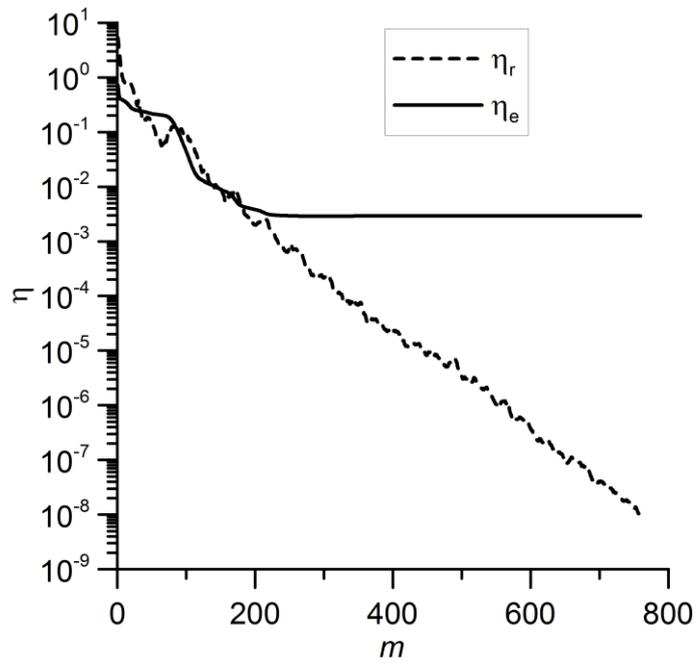

Fig.1. Convergence rate for mesh size $N=41\times41\times41$ and calculation parameters $L_x=1$ cm, $L_y=1$ , cm, $L_z=2$ cm and $L_0=0.8$ cm ($k_0=\pi/L_0$).

While the residual constantly (in average) decreases, the error saturates converging to the error value of the discretization scheme. It is expedient to stop iterations when the error is close to that value. Fig. 2 shows the number of iterations $m_2$ needed to achieve twice less accuracy than the discretization offers as a function of total mesh node numbers $N$ (in this numerical experiment mesh is scaled proportionally).

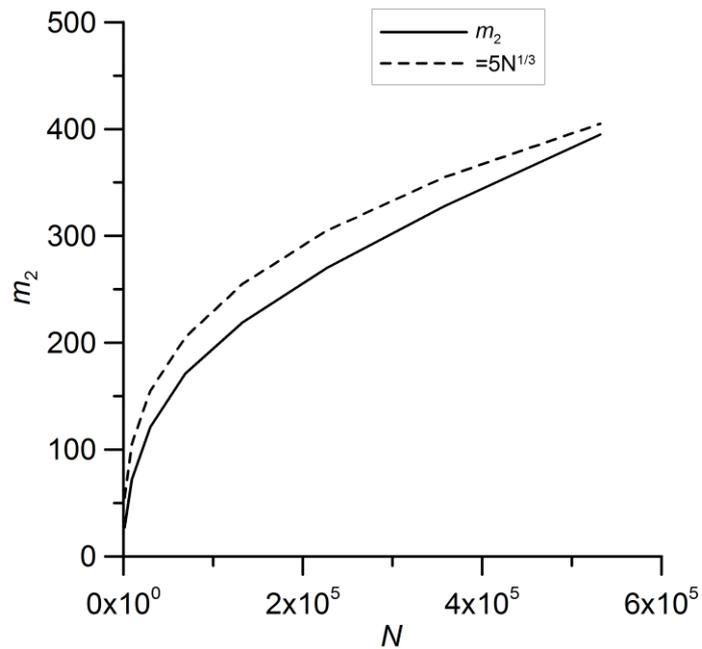

Fig.2. Optimum number of iterations $m_2$ as function of total number of mesh nodes $N$.

So, the optimum number of iterations scales close to $N^{1/3}$ which indicates a good performance of the iteration scheme.

**4. Conclusions**

A new form of time-harmonic Maxwell's equations is developed and proposed for numerical modeling. It is written for the magnetic field strength **H**, electric displacement **D** and vector potential **A** and the scalar potential $\Phi$. There are several attractive features of this form. The first is that the differential operator acting on these quantities is positive. This opens a door for usage, after discretization, of standard iterative procedure such as conjugate gradients. Second is absence of *curl* operators among leading order differential operators. The Laplacian stands for leading order operator in the equations for **H**, **A** and $\Phi$ and the gradient of divergence stands for **D**. This arrangement allows one to use the standard discretization procedure and use the same mesh for all quantities, and also facilitates the boundary condition applications. The third feature is absence of space varied coefficients in the leading order differential operators that provides diagonal domination of the resulting matrix of the discretized equations.

A simple example is given to demonstrate the performance of time-harmonic Maxwell's equations 3D finite difference approach is used with the conjugate gradients method for iterations. Specifics of techniques of handling with boundaries are explained. In the example, quite good conversion of the iterations is achieved.

**Acknowledgements**

This work has been carried out within the framework of the EUROfusion Consortium and has received funding from the Euratom research and training programme 2014-2018 and 2019-2020 under grant agreement No 633053. The views and opinions expressed herein do not necessarily reflect those of the European Commission.

This work also received funding from National Academy of Sciences of Ukraine (grants П-3-22, and ЦВ-5-20).

**References**

1. K. S. Yee, IEEE Trans. Antenna Propagation AP-14 (1966) 302-307. https://doi.org/10.1109/tap.1966.1138693
2. V.E.Moiseenko, O.Agren, J. Plasma Phys 72 (2006) 1133. https://doi.org/10.1017/s0022377806005794

3. V.E.Moiseenko, O. Agren, Phys. Plasmas 12 (2005) 102504. https://doi.org/10.1063/1.2069415
4. V.E.Moiseenko, O.Agren, Second harmonic ion cyclotron heating of sloshing ions in a straight field line mirror, Phys. Plasmas 14 (2007) 022503. https://doi.org/10.1063/1.2435308
5. K. Appert, D. Berger, R. Gruber, and J. Rappaz, J. Comp. Phys. 18 (1975) 284-299. https://doi.org/10.1016/0021-9991(75)90003-0
6. J. C. Nédélec, Numer. Math., 35:3 (1980) 315-341. https://doi.org/10.1007/bf01396415
7. Y.Saad, Iterative Methods for Sparse Linear Systems, PWS Publishing, New York, 2003. https://doi.org/10.1137/1.9780898718003
8. V.E. Moiseenko, Transactions of Fusion Science and Technology 47, №1T (2005) 116-119. https://doi.org/10.13182/fst05-a620
9. Li, D. , Greif, C. and Schötzau, D., Numer. Linear Algebra Appl., 19 (2012) 525-539. https://doi.org/10.1002/nla.782
10. P. Aleynikov and N.B. Marushchenko, Computer Physics Communications 241 (2019) 40-47. https://doi.org/10.1016/j.cpc.2019.03.017